\documentclass[12pt]{article}
\newcommand{\dbox}{\,\raise2pt\hbox{\fbox{\rule{2.5pt}{0pt}\rule{0pt}{2.5pt}}}\,}
\newcommand{\qed}{\,\raise0pt\hbox{\mbox{\rule{6.5pt}{6.5pt}}}}
\newcommand{\bra}[1]{\mbox{$\langle #1 |$}}
\newcommand{\ket}[1]{\mbox{$| #1 \rangle$}}
\newtheorem{lemma}{Lemma}
\newtheorem{theorem}{Theorem}
\begin{document}
\setlength{\baselineskip}{7mm}

\begin{titlepage}
 \begin{normalsize}
  \begin{flushright}
        UT-Komaba/05-9\\
        KEK-TH-1038\\
        September 2005
  \end{flushright}
 \end{normalsize}
 \begin{LARGE}
   \vspace{1cm}
   \begin{center}
       Physical state representations and gauge fixing\\ in string theory\\
   \end{center}
 \end{LARGE}
  \vspace{5mm}
 \begin{center}
    Masako {\sc Asano}%\footnote{\tt asano las osakafu-u ac jp}%
,
    Mitsuhiro {\sc Kato}$^{\dagger}$ %\footnote{\tt kato hep1 c u-tokyo ac jp}
            and
    Makoto {\sc Natsuume}$^{\ddagger}$%\footnote{\tt makoto natsuume kek jp}
\\
      \vspace{4mm}
        {\sl Faculty of Liberal Arts and Sciences}\\
        {\sl Osaka Prefecture University}\\
        {\sl Sakai, Osaka 599-8531, Japan}\\
      \vspace{4mm}
        ${}^{\dagger}${\sl Institute of Physics} \\
        {\sl University of Tokyo, Komaba}\\
        {\sl Meguro-ku, Tokyo 153-8902, Japan}\\
      \vspace{4mm}
        ${}^{\ddagger}${\sl Theory Division,
         Institute of Particle and Nuclear Studies} \\
        {\sl KEK, High Energy Accelerator Research Organization}\\
        {\sl Tsukuba, Ibaraki 305-0801, Japan}\\
      \vspace{1cm}

  %\begin{large} ABSTRACT \end{large}
  ABSTRACT\par
 \end{center}
 \begin{quote}
  \begin{normalsize}
We re-examine physical state representations in the covariant quantization of bosonic string. We especially consider one parameter family of gauge fixing conditions for the residual gauge symmetry due to null states (or BRST exact states), and obtain explicit representations of observable Hilbert space which include those of the DDF states. This analysis is aimed at giving a necessary ingredient for the complete gauge fixing procedures of covariant string field theory such as temporal or light-cone gauge.
\end{normalsize}
 \end{quote}

\end{titlepage}
\vfil\eject

\section{Introduction}

It is well known that null states appearing in the physical Hilbert space of the string theory correspond to the gauge degrees of freedom. For instance, level one null state $L_{-1}|p\rangle$ in the open bosonic string gives the gauge transformation for the massless vector mode on the same level when $p^2=0$. Here $L_n$ ($n$=integer) is the Virasoro operator and $|p\rangle$ is the oscillator vacuum with the momentum eigenvalue $p^{\mu}$ ($\mu=0,1,\cdots,25$). The null states are certainly members of physical states in the sense that they satisfy the physical state condition $L_n|\mbox{phys}\rangle=0$ for positive integer $n$ and the on-shell condition $(L_0-1)|\mbox{phys}\rangle=0$, while they do not contribute to the physical amplitude. In this sense the on-shell physical state is said to have an ambiguity in its representation.

Fixing the gauge degrees of freedom associated with the null state is nothing but taking a representative for the above mentioned ambiguity. One of the well-known representative of physical state is the so-called DDF state~\cite{DelGiudice:1971fp} which is generated by applying the transverse DDF operators to the tachyon state. As will be seen in the subsequent sections, the DDF states are characterized by supplementary linear condition other than the physical state condition.

{}From the point of view of the string field theory (SFT), that extra condition as well as physical state condition plays a role of complete gauge fixing condition of the infinite dimensional gauge symmetry. For the sake of concreteness, let us take the simplest action for the string field $\Psi$ (See e.g., ref.\cite{Banks:1985ff})
\begin{equation}
S = {1\over 2}\Psi(L_0-1)\Psi,\label{SFTaction}
\end{equation}
with the condition
\begin{equation}
L_n\Psi = 0\qquad(n=1,2,\cdots).\label{SFTgauge_cond}
\end{equation}
This is the (partially) gauge-fixed covariant action. Actually these equations lead to the following action and gauge condition for the massless vector field $A^{\mu}$ contained as a mode in $\Psi$
\begin{equation}
S = \int d^{26}x\,\,
{1\over 2}A^{\mu}\dbox A_{\mu},
\end{equation}
\begin{equation}
\partial_{\mu}A^{\mu}=0.
\end{equation}
As is known, the action still has a residual gauge invariance
\begin{equation}
A^{\mu}\rightarrow A^{\mu}+\partial^{\mu}\lambda
\qquad\mbox{with}\qquad\dbox\lambda=0.
\end{equation}
This is also true for the SFT level; eqs.(\ref{SFTaction}) and (\ref{SFTgauge_cond}) has residual gauge invariance
\begin{equation}
\Psi\rightarrow\Psi+L_{-1}\Xi_1+(L_{-2}+{3\over 2}L_{-1}{}^2)\Xi_2,
\end{equation}
provided
\[
\left\{
\begin{array}{rcl}
  L_{n}\Xi_1&=&0\\
  L_{0}\Xi_1&=&0
\end{array}
\right.\qquad\mbox{and}\qquad
\left\{
\begin{array}{rcl}
  L_{n}\Xi_2&=&0\\
  (L_{0}+1)\Xi_2&=&0
\end{array}
\right.
\]
for positive integer $n$.
The degrees of freedom of $\Xi$'s are nothing but the null states (or exact states in the BRST quantization~\cite{Kato:1983im}) mentioned at the beginning. Thus putting some extra conditions, like the DDF representation, corresponds to the complete gauge fixing of SFT.

In the present paper, we will investigate a certain class of complete gauge fixing conditions. In particular, one parameter family of linear gauge condition is analyzed, which is essentially temporal gauge (or chronological gauge) in the sense that the string excitation of a time-like direction is restricted to only zero mode. This family includes the DDF representation as a limit so that the relation between temporal gauge and the light-like gauge will also be clarified.

One of the motivations for studying the temporal gauge and its cousin is that better understanding of the gauge may provide a clue towards the resolution of the long-standing problem on the canonical quantization of SFT~\cite{Eliezer:1989cr}. Since time-like excitation is restricted to the zero mode, it can be taken as a time parameter of canonical quantization procedure and also the interaction becomes local with respect to the time parameter.

For those who are not familiar with the problem may wonder whether there are anything wrong with the SFT because it reproduced the correct quantum amplitudes in perturbative sense. In deriving such amplitudes, however, one assumes that the Feynmann rules can be read off from the action as has been done in the usual local field theories. (See for example ref.\cite{Hata:1986jd}.) There is generally no justification for such an assumption in non-local theories. The existence of light-cone SFT may support the validity of the assumption if the exact relationship from the covariant SFT to the light-cone SFT through the gauge fixing in the SFT level, because in the latter formulation light-like variable $x^+$ is the time parameter of the quantization procedure and locality of the interaction with respect to $x^+$ is satisfied.

In order to try these scenario, as a first step, we will clarify the structure and the representation of the physical states in the temporal gauge in keeping the relation to the DDF states clear, as the latter representation can be regarded as light-like gauge in the SFT level.

This paper is organized as follows. After discussing some generality of gauge fixing and identifying the concrete condition for the DDF states in the next section, we prove in section 3 that a certain class of gauge fixing conditions are complete in the sense that the state space specified by each gauge condition is equivalent to the observable positive definite Hilbert space. It will be also shown there that the representation of observable space given by the DDF states can be obtained by a certain limiting procedure from more general representations, which may cast new light on the relationship between light-like gauge and temporal gauge in the SFT. Section 4 is devoted to the summary and discussions.

\section{Physical states in covariant gauge}

The total state space ${\cal H}(p)$ for the old covariant quantization (OCQ) of perturbative bosonic string theory ($D=26$) is given by the Fock space ${\cal F}\!ock(\alpha_{-n}^\mu ; p^\mu)$ spanned by the states of the form
\begin{equation}
\ket{\phi_N;\,p^\mu} 
= \prod_{\mu=0}^{25} \prod _{n=1}^{\infty} (\alpha_{-n}^{\mu})^{N_n^\mu}
\ket{0;p^\mu} .
\end{equation} 
Here,
$N_n^\mu$ is a non-negative integer and 
 $\ket{0;p^\mu}$ is the ground state annihilated by all $\alpha_{n}^{\mu}$ ($n>0$) with momentum $p^\mu$. We often divide ${\cal H}(p)$ into the space with level $N = \sum_{n,\mu} n N_n^\mu$ as ${\cal H}(p)= \oplus_{N\ge 0} {\cal H}^{(N)}(p) $. 
Among ${\cal H}(p)$, positive definite Hilbert space ${\cal H}_{\rm obs}(p)$ is defined by the quotient ${\cal H}_{\rm obs}(p) = {\cal H}_{\rm phys}(p) / {\cal H}_{\rm null}(p)$, which we sometimes call observable Hilbert space. Here ${\cal H}_{\rm phys}(p)$ is the set of states satisfying the physical state condition 
\begin{equation}
L_n  \ket{\phi}_{\rm phys} =0 \quad (n > 0)
\label{eq:physstcond}
\end{equation}
and the on-shell condition
\begin{equation}
(L_0 - 1) \ket{\phi}_{\rm phys} =0 
\end{equation}
which restricts the level $N$ of the states as $\alpha' p^2 + N -1 = 0$. The space ${\cal H}_{\rm null}(p) [\subset {\cal H}_{\rm phys}(p)] $ is the set of null states that are identified as physical states of the form
\begin{equation}
\ket{\chi}_{\rm null}= L_{-1}\ket{\xi_1} + (L_{-2}+ \frac{3}{2} L_{-1}^2 ) \ket{\xi_2}
\label{eq:nullst}
\end{equation}
where $L_n \ket{\xi_1} = (L_n+\delta_{n,0}) \ket{\xi_2} = 0$ ($ n \ge 0$). A null state has zero inner product with any state in ${\cal H}_{\rm phys}(p)$ 
(${}_{\rm null}\langle \chi \ket{\phi}{}_{\rm phys}=0$). This is seen from (\ref{eq:physstcond}) and (\ref{eq:nullst})
with the definition of inner product in ${\cal H}(p)$: $L_{-n}^\dagger=L_n$ ($\alpha_{-n}^\dagger = \alpha_n $) and $\langle 0;p \ket{0;p}=1$.

Due to the existence of null states, we have an ambiguity $\ket{\phi}_{\rm phys} \sim \ket{\phi}_{\rm phys} + \ket{\chi}_{\rm null}$ in choosing explicit representations of observable Hilbert space ${\cal H}_{\rm obs}(p)$. 
As we have seen in the introduction in terms of SFT,
appearance of null states in our OCQ scheme (or exact states for BRST quantization) indicates the existence of residual gauge symmetry which is left unfixed at the classical level. 
Thus, choosing explicit representation of ${\cal H}_{\rm obs}(p)$ exactly corresponds to fixing this residual gauge symmetry. In fact, in addition to the physical state condition, we need supplementary `gauge condition' which exactly fixes whole gauge degrees of freedom and nothing more nor less:
\begin{equation}
{\cal H}_{\rm phys}(p)\cap\{\mbox{`gauge condition'}\}\sim{\cal H}_{\rm obs}(p).
\label{eq:exobssp}
\end{equation}
We would like to find a class of such conditions 
and corresponding representations of ${\cal H}_{\rm obs}(p)$ 
in a systematic manner. 

For example, we know that the set of so-called DDF states can be taken as an explicit representation of ${\cal H}_{\rm obs}(p)$. They are given by multiplying DDF operators 
\begin{equation}
A_{-n}^{\hat{i}} =
\frac{1}{2\pi} \oint dz \partial X^{\hat{i}}(z) 
e^{-i n k\cdot X(z)}
%\qquad (\hat{i}=1,\cdots, 24)
\end{equation}
on the tachyon ground state $\ket{0,\bar{p}}$ as 
\begin{equation}
\ket{\phi; p=\bar{p}-Nk} =
A_{-n_1}^{\hat{i}_1} A_{-n_2}^{\hat{i}_2}\cdots A_{-n_l}^{\hat{i}_l}
\ket{0,\bar{p}} .
\label{eq:DDFst}
\end{equation}
Here,
\begin{equation}
X^{\mu}=x^{\mu}-ip^{\mu}\ln z 
+ i \sum_{n\ne 0} \frac{1}{n}\alpha_{n}^{\mu}z^{-n},
\end{equation}
$\hat{i}=1,\cdots, 24$, $N= n_1 + \cdots + n_l$, $k^2 =0$ (with $k^{\hat{i}}=0$), $\bar{p}^2=2$ and $\bar{p} \cdot k=1$. Hereafter, we set $\alpha'= 1/2$ ($\alpha_0^\mu = p^{\mu}$). These DDF states satisfy the physical state condition and form a basis of ${\cal H}_{\rm obs}(p)$ for $p=\bar{p}-Nk$. DDF states up to level $N=2$ are as follows:
\begin{eqnarray}
\underline{N=0} : && \ket{0,\bar{p}},
\\
\underline{N=1} : &&  A_{-1}^{\hat{i}} \ket{0,\bar{p}}
 = \alpha_{-1}^{\hat{i}} \ket{0,\bar{p}-k},
\\
\underline{N=2} : &&  
A_{-2}^{\hat{i}} \ket{0,\bar{p}}
 = \left( \alpha_{-2}^{\hat{i}} -2 (k\cdot \alpha_{-1})  \alpha_{-1}^{\hat{i}} \right) 
\ket{0,\bar{p}-2 k},
\label{eq:DDFN2i2}
\\
&& 
A_{-1}^{\hat{i}} A_{-1}^{\hat{j}} \ket{0,\bar{p}}
= 
\left( 
\alpha_{-1}^{\hat{i}} \alpha_{-1}^{\hat{j}} 
+ \frac{1}{2}\delta^{\hat{i}\hat{j}} 
\left[
(k\cdot \alpha_{-1})^2 - (k\cdot \alpha_{-2})   
\right]
\right) 
\ket{0,\bar{p}-2 k}.
\label{eq:DDFN2ij}
\end{eqnarray}
In fact, the set of DDF states can be extracted by imposing an additional condition 
\begin{equation}
k \cdot \alpha_{n} \ket{\phi; \, \bar{p}-Nk} = 0
\quad (n>0)
\label{eq:addcondDDF}
\end{equation}
on the space of physical states ${\cal H}_{\rm phys}(p)$ when $p=\bar{p}-Nk$~\cite{Goddard:1972iy}.
This is an example of supplementary gauge condition that completely fixes the ambiguity of null states as discussed before.
%%%

\section{Representations of observable Hilbert space}

Next we consider a class of supplementary conditions which are linear in oscillator variables. They are simple generalizations of (\ref{eq:addcondDDF}) in the previous section, but still nontrivial in the sense that the proof of (\ref{eq:exobssp}) does not go in the same way as for the DDF states since one cannot define DDF-like operators in general cases.

Concretely, we consider the following condition 
\begin{equation}
\tilde{\epsilon} \cdot \alpha_{n} \ket{\phi; \, p^\mu} = 0
\quad (n>0)
\label{eq:addcond}
\end{equation}
with a constant time-like or light-like vector $\tilde{\epsilon}$
 (i.e., $\tilde{\epsilon}^2\le 0$).
%if we take $\tilde{\epsilon} \cdot p \ne 0$. 
For $\tilde{\epsilon}_\mu \propto k_{\mu}$,
this condition reduces to (\ref{eq:addcondDDF}) and restricts states not to include $\alpha_{-n}^{-} = ( \alpha_{-n}^{0} - \alpha_{-n}^{25}) / \sqrt{2}$ or $\alpha_{-n}^{+} = ( \alpha_{-n}^{0} + \alpha_{-n}^{25}) / \sqrt{2}$ respectively for 
$ k_\mu \propto \delta_{\mu}{}^0 + \delta_{\mu}{}^{25}$ or  
$k_\mu \propto \delta_{\mu}{}^0 - \delta_{\mu}{}^{25}$. 
Also, for $\tilde{\epsilon}_\mu = \delta_{\mu}{}^0$, the condition (\ref{eq:addcond}) restricts states not to include any time-like oscillators ($\alpha_{-n}^0$). 
We do not consider the case $\tilde{\epsilon}^2 > 0$ since 
the condition for such a case is not practical as a gauge condition,
though the condition itself works well to satisfy (\ref{eq:exobssp}) with some appropriate assumptions.

\subsection{Main theorem}

The main claim of the present paper is the following theorem:
\begin{theorem}
Let ${\cal H}_{\tilde{\epsilon}}(p)$ denotes 
the subspace of ${\cal H}(p)$ spanned by the states
satisfying both 
$$
(L_n - \delta_{n,0}) \ket{\phi}_{\rm phys} =0 \quad (n\ge 0)
$$
and 
$$
\tilde{\epsilon} \cdot \alpha_{n} \ket{\phi; \, p^\mu} = 0
\quad (n>0)
$$
for  $\tilde{\epsilon}^2 \le 0$. 
Then, ${\cal H}_{\tilde{\epsilon}}(p) \sim {\cal H}_{\rm obs}(p)$ 
provided $\tilde{\epsilon} \cdot p \ne 0$.
\label{thm1}
\end{theorem}
%%%%%%
We divide ${\cal H}_{\tilde{\epsilon}}(p)$ by level $N$ as ${\cal H}_{\tilde{\epsilon}}(p)=\oplus_{N\ge 0}  {\cal H}_{\tilde{\epsilon}}^{(N)}(p)$ and 
prove the theorem for each $N$. 
Before going into general proof, let us first see the simple cases $N=0$ and $N=1$.
For $N=0$, we only have ground state $\ket{0,p}$ in ${\cal H}^{(0)}(p)$ (with $p^2=2$) and it satisfies (\ref{eq:physstcond}) and (\ref{eq:addcond}) trivially: 
${\cal H}_{\tilde{\epsilon}}^{(0)}(p)=\{ \ket{0,p} \}\,(= {\cal H}_{\rm obs}^{(0)}(p))$.
For $N=1$, general on-shell states satisfying (\ref{eq:physstcond}) are represented as 
\begin{equation}
\ket{\phi;\,p} = \xi \cdot \alpha_{-1} \ket{0,p}
\end{equation}
with $p^2=0$ and $\xi \cdot p =0$. Among these states, there is a null state $p \cdot \alpha_{-1} \ket{0,p}$ and the space ${\cal H}^{(1)}_{\rm obs}(p)$ is identified up to the ambiguity $\xi_\mu \sim \xi_{\mu}+ p_\mu$. The condition (\ref{eq:addcond}) gives the constraint on $\xi_\mu $ as $\tilde{\epsilon} \cdot \xi=0$, which fixes the ambiguity completely since $\tilde{\epsilon} \cdot p \ne 0$ is assumed. 
Thus, ${\cal H}^{(1)}_{\tilde{\epsilon}}(p) \sim {\cal H}^{(1)}_{\rm obs}(p)$. 
Explicitly, ${\cal H}_{\tilde{\epsilon}}^{(1)}(p)=
\{\xi \cdot \alpha_{-1}  \ket{0,p}\;|\; p \cdot \xi = \tilde{\epsilon}\cdot \xi =0\}$
for on-shell $p$ ($p^2=0$) with $\tilde{\epsilon} \cdot p \ne 0$.
We have proven ${\cal H}_{\tilde{\epsilon}}(p)\sim {\cal H}_{\rm obs}(p) $ for $N=0, 1$. Note that here we have not used the condition $\tilde{\epsilon}^2 \le 0$. 
%Indeed, the theorem is satisfied for any constant vector $\tilde{\epsilon}_{\mu}$ with $\tilde{\epsilon} \cdot p \ne 0$ for $N=0,1$. 
For general $N$, we first give a proof for $\tilde{\epsilon}^2 < 0$ and then extend it to $\tilde{\epsilon}^2=0$ since the latter can be considered as a limit of the former. For $N\ge 2$, the condition $\tilde{\epsilon} \cdot p \ne 0$ is always satisfied for on-shell states if $\tilde{\epsilon}^2 \le 0$. 

%%%%%
\subsection{Proof for {\boldmath $\tilde{\epsilon}^2 < 0 $} } 
First, we will make some definitions for preparation%
\footnote{In fact, to prove the theorem~\ref{thm1} for $\tilde{\epsilon}^2 < 0$, it is sufficient to take $\tilde{\epsilon}_\mu=\delta_{\mu}^{0} $
since other cases can be obtained by boost transformations from this. 
We however consider every $\tilde{\epsilon}^2 < 0$ explicitly for later convenience.
}.
We will fix the time-like vector $\tilde{\epsilon}_{\mu}$ as
\begin{equation}
\tilde{\epsilon}_{\mu} = (\cosh\beta, 0, \cdots, 0, \sinh\beta)\; 
\left[ = \tilde{\epsilon}_{\mu}(\beta) \right]
\end{equation}
with $0\le \beta < \infty$ without losing generality.
Correspondingly, we define a space-like vector
\begin{equation}
   {\epsilon}_{\mu}(\beta) = (\sinh\beta, 0, \cdots, 0, \cosh\beta) .
\end{equation}
We take a particular choice of spacetime coordinates $(t_\beta, s_\beta,x^{\hat{i}}) \equiv
 (\tilde{\epsilon}(\beta) \cdot x, \epsilon(\beta) \cdot x , x^{\hat{i}})$ which are obtained by boost transformation from the original coordinates $x^{\mu}$.
Commutation relations for $\alpha_{n}^{\tilde{\mu}}$ ($\tilde{\mu} = t_\beta, s_\beta, \hat{i}$) are given as
\begin{equation}
 [ \alpha_m^{\tilde{\mu}}, \alpha_n^{\tilde{\nu}}] 
= m \delta_{m+n,0} \eta^{\tilde{\mu} \tilde{\nu}} 
\label{eq:commrel}
\end{equation}
where 
\begin{eqnarray}
 \alpha_{n}^{t_{\beta}} &=& \cosh\beta \, \alpha_{n}^{0} 
+ \sinh\beta \, \alpha_{n}^{25} \; (= \tilde{\epsilon}({\beta}) \cdot \alpha_{n}),
\\
\alpha_{n}^{s_{\beta}} &=& \sinh\beta \, \alpha_{n}^{0} 
+ \cosh\beta \, \alpha_{n}^{25} \; (= {\epsilon}({\beta}) \cdot \alpha_{n}).
\end{eqnarray}
Thus, in particular,
\begin{equation}
[
\tilde{\epsilon}({\beta}) \cdot \alpha_{m},
\tilde{\epsilon}({\beta}) \cdot \alpha_{n}
]
= -m  \delta_{m+n,0} 
,\quad
[
{\epsilon}({\beta}) \cdot \alpha_{m},
{\epsilon}({\beta}) \cdot \alpha_{n}
]
=  m  \delta_{m+n,0} 
,\quad
[\tilde{\epsilon}({\beta}) \cdot \alpha_{m},
{\epsilon}({\beta}) \cdot \alpha_{n} ]=0
.
\end{equation}
We divide total state space into `time-like'  and `space-like' part: 
\begin{equation}
{\cal H}(p) = {\cal H}_{t_\beta}(p^{t_{\beta}}) 
\otimes {\cal H}_{\Sigma_\beta}({p^{i_\beta}}) 
\end{equation}
where $i_\beta=(s_{\beta},\hat{i})$, 
${\cal H}_{t_\beta}(p^{t_{\beta}}) 
= {\cal F}\!ock(\alpha_{-n}^{t_{\beta}} ;\, p^{t_{\beta}})$
and 
${\cal H}_{\Sigma_\beta}({p^{i_\beta}}) =
{\cal F}\!ock(\alpha_{-n}^{i_\beta} ;\, p^{i_\beta})$. 
We also divide $L_n$ as $L_n = L_n^{t_\beta} + L_n^{\Sigma_\beta}$ 
where 
\begin{equation} 
L_n^{t_\beta} = - \frac{1}{2} \sum_{m=-\infty}^{\infty}
: \alpha_{n-m}^{t_\beta} \alpha_{m}^{t_\beta} : \; ,
\quad
L_n^{\Sigma_\beta} =
\frac{1}{2} \sum_{m=-\infty}^{\infty}
: \alpha_{n-m}^{i_\beta} \alpha_{m}^{i_\beta} : \,.
\end{equation}
We further define the space ${\cal F}_{\beta}(p)$ as 
\begin{equation}
{\cal F}_{\beta}(p)=\{
\ket{f_{\beta};\,p } \;|\;\,
\alpha_n^{t_\beta} \ket{f_{\beta};\,p } = L_n \ket{f_{\beta};\,p } =0 
\;\, (n>0) 
\}.
\end{equation}
The relation between this ${\cal F}_{\beta}(p)$ 
and ${\cal H}_{\tilde{\epsilon}(\beta)}(p)$ is 
\begin{equation}
{\cal H}_{\tilde{\epsilon}(\beta)}(p) = 
\{ 
\ket{\phi}\in {\cal F}_{\beta}(p)\,|\, (L_0-1) \ket{\phi}=0
\} .
\end{equation}
The space ${\cal F}^{(N)}_{\beta}(p)$ 
is a subspace of $\ket{0,p^{t_\beta}} \otimes {\cal H}^{(N)}_{\Sigma_\beta}({p^{i_\beta}})$ since $\ket{f_{\beta};\,p }$ does not contain any `time-like' oscillator $\alpha_{-n}^{t_\beta}$.
Thus, ${\cal F}_{\beta}(p)$ is positive definite and cannot contain null states (\ref{eq:nullst}).

With the above definitions, we will now begin to prove theorem~1, i.e., 
$ {\cal H}^{(N)}_{\tilde{\epsilon}(\beta)}(p)
\sim {\cal H}^{(N)}_{\rm obs}(p)$. 
%for $N=-\frac{1}{2} p^2 + 1$. 
First, we will give the following lemma:
%%%
\begin{lemma}
States of the form 
\begin{equation}
L_{-n_1} \cdots L_{-n_r}
L^{t_\beta}_{-m_1} \cdots L^{t_\beta}_{-m_q} 
\ket{f_{\beta};\,p },
\qquad
\ket{f_{\beta};\,p } \in {\cal F}_{\beta}(p)
\label{eq:statebase1}
\end{equation}
($n_s \le n_{s+1}, m_s \le m_{s+1}$) are linearly independent and span a basis of ${\cal H}(p)$  
if $p^{t_\beta}\ne 0$.
\label{lemma1}
\end{lemma}
%%%
The proof is given in Appendix~A.
\bigskip

With the above lemma, we will write every state in ${\cal H}(p)$ as a sum of states of the form (\ref{eq:statebase1}). 
In particular, we divide any $\ket{\mbox{phys}} \in {\cal H}_{\rm phys}(p)$ written in this form into two classes as  
\begin{equation}
\ket{\mbox{phys}} = \ket{g} + \ket{\chi} 
\end{equation} 
where $\ket{g}$ consists of terms without any $L_{-n}$, i.e., 
\begin{equation}
\ket{g} = 
\sum C_{m_1,\cdots, m_q} L^{t_\beta}_{-m_1} \cdots L^{t_\beta}_{-m_q} 
\ket{f_{\beta};\,p }
\end{equation}
with constants $C_{m_1,\cdots, m_q}$ and the $\ket{\chi} $ part consists of terms including at least one $L_{-n}$. 
Both $\ket{g}$ and $\ket{\chi}$ satisfy on-shell condition.
Also, we see that $L_1 \ket{g}$ and $(L_2+\frac{3}{2}L_1^2) \ket{g}$ do not contain any $L_{-n}$ and $L_1 \ket{\chi} $ and $(L_2+\frac{3}{2}L_1^2) \ket{\chi} $ again consist of terms with at least one $L_{-n}$. 
Thus $L_n \ket{\mbox{phys}} = 0$ implies 
that $\ket{g}$ and $\ket{\chi} $ are both physical and the state $\ket{\chi} $ is null since all $L_{n}$ $(n\ge 1)$ are generated by $L_{1}$ and $L_{2}$.
For $\ket{g}$ part, 
$
0= (L_n^{t_\beta}+ L_n^{\Sigma_\beta} ) \ket{g} 
= L_n^{t_\beta} \ket{g} 
$
for any $n > 0$ since $L_n^{\Sigma_\beta}\ket{f_{\beta};\,p }=0$. 
This contradicts the non-degeneracy of $c=1$ Verma module% 
\footnote{
${\cal V}(c,h)$ is a linear space spanned by the states 
constructed by acting Virasoro operators  ($L_{-n}$, $n>0$) 
of central charge $c$ on the highest weight state $\ket{h}$. 
}
${\cal V}(1,h<0)$ if there exist any $L^{t_\beta}_{-m}$ in $\ket{g}$.
This means that $\ket{g}$ contains no $L^{t_\beta}_{-m}$
and 
\begin{equation}
\ket{g} =  \ket{f_{\beta};\,p } .
\end{equation}
Thus we have shown that any physical state can be written as an element of 
${\cal H}_{\tilde{\epsilon}(\beta)}(p) \subset  {\cal F}_{\beta}(p) $ up to a null state: 
\begin{equation}
\ket{\mbox{phys}} = \ket{f_{\beta};\,p }  + \ket{\chi} ,
\quad
\ket{f_{\beta};\,p } \in {\cal H}_{\tilde{\epsilon}(\beta)}(p) . 
\end{equation}
In other word, we have shown that $
{\cal H}_{\tilde{\epsilon}(\beta)}(p)
\sim {\cal H}_{\rm obs}(p)$ since we know that there are no null states in 
${\cal H}_{\tilde{\epsilon}(\beta)}(p)$. We have proven theorem~\ref{thm1} for $\tilde{\epsilon}^2<0$. \qed

\bigskip

Note that in some parts of the above proof 
we have used the similar 
argument given in ref.\cite{Goddard:1972iy,Thorn:1986xy} 
where essentially the same statement as our theorem~\ref{thm1} 
for the set of DDF operators (which corresponds to our case of $\tilde{\epsilon}^2 = 0$ and  $p^\mu = \bar{p}^\mu - Nk^\mu$)
has been proved.
Comparing to that case, our proof for $\tilde{\epsilon}^2 < 0$
is rather simpler since the positive-definiteness of 
${\cal F}_{\beta}(p)$ is trivial (and also we know the non-degeneracy of ${\cal V}(1,h<0)$). 
%due to Kac determinant.

%%%%%%%%%%%%%%%%%%%%%%%%%%%%%%%%%%%%%%%%%%%%%%%
\subsection{{\boldmath Properties of $ {\cal H}_{\tilde{\epsilon}(\beta)}(p) $}}
In this subsection, we present some properties of the space 
${\cal H}_{\tilde{\epsilon}(\beta)}(p) $ as a representative of  observable Hilbert space ${\cal H}_{\rm obs}(p)$. 

The dimension of ${\cal H}_{\rm obs}(p)$ coincides with that of 
the transverse Hilbert space 
${\cal H}(p^{\hat{i}})= {\cal F}\!ock(\alpha_{-n}^{\hat{i}};p^{\hat{i}}) $:
For each level $N$, $\dim {\cal H}^{(N)}_{\rm obs}(p) = P_{24}(N)$
where $P_{D}(n)$ is the coefficient of $q^{n}$ in $\prod_{n\ge 1}(1-q^n)^{-D}$.
We would like to choose a basis of ${\cal H}_{\tilde{\epsilon}(\beta)}(p) $
in order to analyze the space systematically.
For this aim, we have the following lemma~\cite{ip}
\begin{lemma}
Assume that $p^{t_\beta} (=\tilde{\epsilon} \cdot p) \ne 0$ and 
$p^{s_\beta} (=\epsilon \cdot p) \ne \frac{r-s}{\sqrt{2}}$
where $r$ and $s$ are positive integers with $rs<N$.
Then, a state $\ket{f_{\beta};\,p } \in {\cal H}^{(N)}_{\tilde{\epsilon}(\beta)}(p) $ has at least one term consisting only of transverse oscillators, i.e.,
\begin{equation}
\ket{f_{\beta};\,p } = \ket{ \hat{\phi} ;\,p }
+ (\mbox{terms with at least one $\alpha_{-n}^{s_\beta}$}) 
\end{equation}
where $\ket{\hat{\phi} ;\,p }$ is a non-zero state in ${\cal F}\!ock(\alpha_{-n}^{\hat{i}};p)$.
\end{lemma}
With this result, for $p^{s_\beta} \ne \frac{r-s}{\sqrt{2}}$, we can choose a basis of 
${\cal H}^{(N)}_{\tilde{\epsilon}(\beta)}(p) $ as follows: 
We specify each basis element $\ket{f_{\beta};\,p }_{\hat{\lambda}_N}$
of ${\cal H}^{(N)}_{\tilde{\epsilon}(\beta)}(p) $ by the term 
\begin{equation}
\ket{\hat{\phi} ;\,p}_{\hat{\lambda}_N = \{(\hat{i}_1,n_1),\cdots,(\hat{i}_l,n_l)\}} = 
\alpha_{-n_1}^{\hat{i}_1} \cdots \alpha_{-n_l}^{\hat{i}_l} \ket{0,\;p},
\quad (n_s \le n_{s+1}, \sum_{s=1}^l n_s = N)
\label{eq:flambdai}
\end{equation}
and write 
\begin{equation}
\ket{f_{\beta};\,p }_{\hat{\lambda}_N} = \ket{\hat{\phi} ;\,p}_{\hat{\lambda}_N}  
+ (\mbox{terms with at least one $\alpha_{-n}^{s_\beta}$}).
\end{equation}
With fixed $\hat{\lambda}_N$,
the terms with $\alpha_{-n}^{s_\beta}$ in $\ket{f_{\beta};\,p }_{\hat{\lambda}_N}$
are uniquely determined by the condition $L_n  \ket{f_{\beta};\,p }_{\hat{\lambda}_N}=0$.
Note that for $|p^{s_\beta}| > \frac{N-1}{\sqrt{2}}$ we can always choose the above basis 
since $p^{s_\beta} \ne \frac{r-s}{\sqrt{2}}$ and $p^{t_\beta} \ne 0$ for such a case.

For example, we explicitly represent the space ${\cal H}^{(N)}_{\tilde{\epsilon}(\beta)}(p) $ for $N=1,2$ 
by the basis given above.
For $N=1$ with $p^{s_\beta}\ne 0$, 
\begin{equation}
{\cal H}^{(N=1)}_{\tilde{\epsilon}(\beta)}(p) 
=
\{ \ket{f_{\beta};\,p }_{\hat{\lambda}_1=(\hat{i},1)  } \}
\end{equation}
where
\begin{equation}
\ket{f_{\beta};\,p }_{(\hat{i},1)}
= \left[\alpha_{-1}^{\hat{i}} - p^{\hat{i}}\,
\frac{\alpha^{s_\beta}_{-1} }{p^{s_\beta}}
%\frac{\epsilon \cdot \alpha_{-1} }{\epsilon \cdot p}
\right]\, \ket{0,p}.
\label{eq:N1i}
\end{equation}
For $N=2$ with $p^{s_\beta}\ne 0, \pm\frac{1}{\sqrt{2}}$,
\begin{equation}
{\cal H}^{(N=2)}_{\tilde{\epsilon}(\beta)}(p) 
=
\{ 
\ket{f_{\beta};\,p }_{\hat{\lambda}_2= \{(\hat{i},2) \}} \;, 
\ket{f_{\beta};\,p }_{\hat{\lambda}_2=  \{(\hat{i},1),(\hat{j},1) \} }
 \}
\end{equation}
where
\begin{equation}
\ket{f_{\beta};\,p }_{(\hat{i},2)} = 
\left[
\alpha_{-2}^{\hat{i}} 
-\frac{2}{p_{s_\beta}} \alpha_{-1}^{\hat{i}} \alpha_{-1}^{s_\beta}
+ \frac{ 4 p^{\hat{i}} } {2 p_{s_\beta}^2-1}
 \alpha_{-1}^{s_\beta} \alpha_{-1}^{s_\beta}
- \frac{p^{\hat{i}}(2 p_{s_\beta}^2 +1)}{ p_{s_\beta}(2 p_{s_\beta}^2-1) } \alpha_{-2}^{s_\beta}
\right]  \ket{0,p}
\label{eq:N2i2}
\end{equation}
and
\begin{equation}
\ket{f_{\beta};\,p }_{\{(\hat{i},1),(\hat{j},1)\}} = 
\left[
\alpha_{-1}^{\hat{i}} \alpha_{-1}^{\hat{j}} 
-\frac{2}{p_{s_\beta}} p^{\{ \hat{i}}\alpha_{-1}^{\hat{j} \} } \alpha_{-1}^{s_\beta}
+\frac{ \delta^{\hat{i}\hat{j}} +2 p^{\hat{i}} p^{\hat{j}}} {2 p_{s_\beta}^2-1}
 \alpha_{-1}^{s_\beta} \alpha_{-1}^{s_\beta}
- \frac{p_{s_\beta}^2 \delta^{\hat{i}\hat{j}} + p^{\hat{i}} p^{\hat{j}}}
 {p_{s_\beta}(2 p_{s_\beta}^2-1)} \alpha_{-2}^{s_\beta}
\right]  \ket{0,p}.
\label{eq:N2ij}
\end{equation}

%%%%%%%%%%%%%%%%%%%%%%%%%%%%%%%%%%%%%%%%%%%%%%%
\subsection{{\boldmath Proof for $\tilde{\epsilon}^2 = 0 $}}

Now we prove the theorem for 
the remaining case: $\tilde{\epsilon}^2 = 0$.
In this case, we may say that we already have a proof in ref.\cite{Goddard:1972iy,Thorn:1986xy}. 
We will however give a proof based on the new picture 
where the representation of physical states 
${\cal H}_{\tilde{\epsilon}}(p)$ for $\tilde{\epsilon}^2 = 0$ 
can be understood as a limit of that  for $\tilde{\epsilon}^2 < 0$. 
In other word, we will identify the space 
${\cal H}_{\tilde{\epsilon}(\beta=\infty)}(p)$ as a limit 
`$\lim_{\beta\to \infty} 
{\cal H}_{\tilde{\epsilon}(\beta)}(p) 
$.' 
In order to define such a limit consistently, 
we choose a set of particular states as a basis of space ${\cal H}_{\tilde{\epsilon}(\beta)}(p) $
and take the limit%
\footnote{The limit we consider is different from boost transformation since we 
keep the momentum $p^\mu$ fixed.
}
for each basis element of the space for fixed momentum~$p^\mu$.

Now, we will explain how to define the limit explicitly.
We consider the space ${\cal H}^{(N)}_{\tilde{\epsilon}(\beta)}(p) $ with fixed on-shell momentum $p^\mu$ for each $N$.
Here the momentum frame has to be chosen in order to satisfy
$\tilde{\epsilon}(\beta) \cdot p \ne 0$ for arbitrary $\beta \,(\le \infty)$, i.e.,
we take $p^0+p^{25} \ne 0$. 
Then we take $\beta$ large enough ($\beta > \beta_0^{N}$)
to satisfy $|\epsilon(\beta) \cdot p| >\frac{N-1}{\sqrt{2}} $ for 
each $p^{\mu}$ and $N$.
We can always take such $\beta_0^{N}$ since $\lim_{\beta\to \infty} |\epsilon(\beta) \cdot p| =\infty$ for any $p$ with $p^0+p^{25} \ne 0$.
{}From the discussion of the previous subsection, we can take the set of states $\{ \ket{f_{\beta};\,p }_{\hat{\lambda}_N}  \}$ as a basis of ${\cal H}^{(N)}_{\tilde{\epsilon}(\beta)}(p) $
for $\beta > \beta_0^{N}$.
Each state $\ket{f_{\beta};\,p }_{\hat{\lambda}_N}$  
contains $\beta$ through the parts of ${\epsilon}(\beta) \cdot \alpha_{-n}$ ($n\ge 0$) 
and thus the state can be expanded with respect to $e^{\beta}$.
We can prove from the property of physical state condition 
that the terms with positive powers of $e^{\beta}$ 
cannot appear in the expansion of $\ket{f_{\beta};\,p }_{\hat{\lambda}_N}$  and 
\begin{equation}
 \lim_{\beta(>\beta_0^{N})\to \infty} \ket{f_{\beta};\,p }_{\hat{\lambda}_N} < \infty. 
\label{eq:limitfbeta}
\end{equation}
Also, the terms with odd powers of $e^{\beta}$ do not appear in the expansion and 
thus the expansion takes the form 
\begin{eqnarray}
\ket{f_{\beta} ;\,p }_{\hat{\lambda}_N}
&=&
\ket{f^{(0)};\,p }_{\hat{\lambda}_N} + e^{-2\beta} \ket{f^{(1)};\,p }_{\hat{\lambda}_N} 
+ e^{-4\beta} \ket{f^{(2)};\,p }_{\hat{\lambda}_N} + \cdots
\nonumber\\
&=& \sum_{k=0}^{\infty}
e^{-2k\beta} \ket{f^{(k)};\,p }_{\hat{\lambda}_N}. 
\end{eqnarray}
The leading term $\ket{f^{(0)};\,p }_{\hat{\lambda}_N}$ is given by the limit 
(\ref{eq:limitfbeta}) and contains the term 
$\ket{\hat{\phi} ;\,p}_{\hat{\lambda}_N}$ of (\ref{eq:flambdai}).
By definition, each term $\ket{f^{(k)} ;\,p }_{\hat{\lambda}_N}$ does not contain 
$\beta$ and satisfies physical state condition 
\begin{equation}
L_n \ket{f^{(k)} ;\,p }_{\hat{\lambda}_N}=0. 
\label{eq:limphyscond}
\end{equation}
Furthermore, from the condition 
$\tilde{\epsilon}(\beta) \cdot \alpha_{n} \ket{f_{\beta} ;\,p }_{\hat{\lambda}_N}=0$,
we have
\begin{equation}
(\alpha_n^{0} + \alpha_n^{25}) \ket{f^{(k)};\,p}_{\hat{\lambda}_N} 
+ (\alpha_n^{0} - \alpha_n^{25}) \ket{f^{(k-1)};\,p}_{\hat{\lambda}_N} =0.
\end{equation}
In particular, the leading term 
$ \ket{f^{(0)};\,p }_{\hat{\lambda}_N} \, (=\lim_{\beta\to \infty} \ket{f_{\beta} ;\,p }_{\hat{\lambda}_N} )$
satisfies 
\begin{equation}
(\alpha_n^{0} + \alpha_n^{25})  \ket{f^{(0)};\,p }_{\hat{\lambda}_N}  \,
(\propto \tilde{\epsilon}{(\beta\to \infty)} 
\cdot \alpha_n \ket{f^{(0)};\,p }   ) \,=0.
\label{eq:limgaugecond}
\end{equation}

The limit of the inner product of two states 
$\ket{f_{\beta} ;\,p }_{\hat{\lambda}_N}$ and $\ket{f_{\beta} ;\,p }_{\hat{\lambda}_N'}$ 
can be explicitly calculated as
\begin{eqnarray}
&& \lim_{\beta \to \infty}\;
{}_{\hat{\lambda}_N} \langle f_{\beta} ;\,p \,\ket{f_{\beta} ;\,p }_{\hat{\lambda}_N'} 
\;(= 
{}_{\hat{\lambda}_N} \langle f^{(0)};\,p \,  \ket{f^{(0)};\,p }_{\hat{\lambda}_N'}   )
\nonumber\\
& & \hspace*{2cm}=
{}_{\hat{\lambda}_N} \langle \hat{\phi} ;\,p \,
\ket{\hat{\phi} ;\,p}_{\hat{\lambda}_N'}
\nonumber\\
& & \hspace*{2cm}=
f_{\hat{\lambda}_N} \, \delta_{\hat{\lambda}_N, \hat{\lambda}_N'}
\end{eqnarray}
where $f_{\hat{\lambda}_N}$ is a positive integer.
This means that the space spanned by the states
$\lim_{\beta\to\infty} \ket{f_{\beta} ;\,p }_{\hat{\lambda}_N} $
with all $\hat{\lambda}_N$ has dimension $P_{24}(N)$ and is non-degenerate.
Thus, 
\begin{equation}
\{\lim_{\beta\to\infty} \ket{f_{\beta} ;\,p }_{\hat{\lambda}_N} \}
= {\cal H}^{(N)}_{\tilde{\epsilon}(\beta=\infty)}(p) 
\sim
{\cal H}^{(N)}_{\rm obs}(p)
\end{equation}
from (\ref{eq:limphyscond}) and (\ref{eq:limgaugecond}).
We have proven theorem~1 for $\tilde{\epsilon}^2=0$. \qed

\bigskip

The characteristic point of our proof 
comparing to the one in the literature \cite{Goddard:1972iy,Thorn:1986xy}
is that the non-degeneracy of the space ${\cal H}_{\tilde{\epsilon}(\beta=\infty)}(p) 
$
is easily seen from that of $ {\cal H}_{\tilde{\epsilon}(\beta<\infty)}(p) 
$ and 
each state in ${\cal H}_{\tilde{\epsilon}(\beta=\infty)}(p) $ is represented as a limit of the 
corresponding state in ${\cal H}_{\tilde{\epsilon}(\beta<\infty)}(p) 
$.
In fact,  the space ${\cal H}_{\tilde{\epsilon}(\beta=\infty)}(p) $ 
coincides with a set of DDF states if $p^{\hat{i}} = 0$. Explicitly, 
\begin{equation}
\ket{f^{(0)};\,p }_{\hat{\lambda}_N= \{(\hat{i}_1,n_1),\cdots,(\hat{i}_l,n_l)\}} 
= A_{-n_1}^{\hat{i}_1} \cdots A_{-n_l}^{\hat{i}_l} \ket{0,p + N k}
\end{equation}
where $k$ is a light-like vector defined by $k \propto \lim_{\beta\to \infty }\tilde{\epsilon}(\beta)$ (i.e., $k_{\mu} \propto (1, 0,\cdots,0,1)$) and $k\cdot p =1$.

For example, for $N=1$, we explicitly take the $\beta \to \infty$ limit of 
(\ref{eq:N1i}):
By using
\begin{equation}
\lim_{\beta\to \infty} \frac{\alpha_{-n}^{s_\beta}}{p^{s_\beta}}
= \lim_{\beta\to \infty} \frac{\epsilon(\beta)\cdot \alpha_{-n}}
{\epsilon(\beta) \cdot p}
=k \cdot  \alpha_{-n},
\end{equation}
we obtain 
\begin{equation}
\lim_{\beta\to\infty} \ket{f_{\beta};\,p }_{(\hat{i},1)}
= 
\left(
\alpha_{-1}^{\hat{i}}  
-  
p^{\hat{i}} (k \cdot \alpha_{-1} ) 
\right)
\ket{0,p}
\end{equation}
and this coincides with DDF state $A_{-1}^{\hat{i}} \ket{0,p+k}$ if we take $p^{\hat{i}} = 0$.
For $N=2$, we can similarly take the limit of 
(\ref{eq:N2i2}) and (\ref{eq:N2ij})
and the result for $p^{\hat{i}}=0$ is 
\begin{equation}
\lim_{\beta\to\infty} \ket{f_{\beta};\,p }_{(\hat{i},2)}
= 
\left(
\alpha_{-2}^{\hat{i}}  
-  2
 (k \cdot \alpha_{-1} ) \alpha_{-1}^{\hat{i}}
\right)
\ket{0,p}
\end{equation}
and
\begin{equation}
\lim_{\beta\to\infty} 
\ket{f_{\beta};\,p }_{\{(\hat{i},1),(\hat{j},1)\}} = 
\left(
\alpha_{-1}^{\hat{i}} \alpha_{-1}^{\hat{j}} 
+ \frac{1}{2} \delta^{\hat{i}\hat{j}} 
\left[
(k \cdot \alpha_{-1}  )^2 
- (k \cdot \alpha_{-2} )
\right]
\right)
\ket{0,p}\, ,
\end{equation}
which coincide with DDF states (\ref{eq:DDFN2i2}) and (\ref{eq:DDFN2ij}).

%%%%%
\section{Summary and Discussions}

In the present paper, 
we have investigated the old covariant quantization of bosonic string theory and identified a class of additional conditions which precisely fix the residual gauge symmetry corresponding to the ambiguity of null states.
By imposing such an additional condition on the space of physical states, 
we obtain a space which can be taken as an explicit representation of observable  Hilbert space ${\cal H}_{\rm obs}(p)$.
Explicitly, we have proven that the condition $\tilde{\epsilon} \cdot \alpha_{n} \ket{\phi; \, p^\mu} = 0$ for a constant time-like or light-like $\tilde{\epsilon}$ exactly plays the role of the additional gauge condition which precisely fix the ambiguity of null states if 
$\tilde{\epsilon}$ is chosen as $\tilde{\epsilon} \cdot p \ne 0$.
As a result, for each $\tilde{\epsilon}$, we have identified the space ${\cal H}_{\tilde{\epsilon}}(p)$ which gives a complete set of physical states as a particular representation of ${\cal H}_{\rm obs}(p)$.

For time-like $\tilde{\epsilon}=\tilde{\epsilon}(\beta\!<\!\infty)$, the additional condition is related to the temporal gauge in the sense that the corresponding representation of observable Hilbert space ${\cal H}_{\tilde{\epsilon}(\beta<\infty)}(p)$ does not include time-like oscillators $\alpha_{-n}^{t_\beta}$. 
On the other hand, the condition for light-like $\tilde{\epsilon}=\tilde{\epsilon}(\beta\!=\!\infty)$ is related to the light-cone gauge and in this case the space ${\cal H}_{\tilde{\epsilon}(\beta=\infty)}(p)$ consists of physical states without $\alpha_{-n}^{-}$.
For each case, we have also identified a particular basis of ${\cal H}_{\tilde{\epsilon}}(p)$, which would be useful for analyzing the theory (especially SFT) in the corresponding gauge. 
In particular, the space ${\cal H}_{\tilde{\epsilon}(\beta=\infty)}(p)$ for $p^{\hat{i}}=0$ is equivalent to the set of DDF states.
As for the other cases, our result means that we have systematically obtained a class of complete sets of physical states other than the DDF states.
We have also seen that the bases we used for ${\cal H}_{\tilde{\epsilon}(\beta<\infty)}(p)$ and for ${\cal H}_{\tilde{\epsilon}(\beta=\infty)}(p)$ are in one-to-one correspondence, i.e., we have shown that each state in ${\cal H}_{\tilde{\epsilon}(\beta=\infty)}(p) $ (for $p^{+} \ne 0$) is obtained as a certain limit of the corresponding state in ${\cal H}_{\tilde{\epsilon}(\beta<\infty)}(p)$ except for a particular value of momentum vector.
This means that there is a close relation between those two types of representations of physical states and it might be possible that there is a substantial structure for the states in ${\cal H}_{\tilde{\epsilon}(\beta<\infty)}(p)$ as well as for DDF states.
Further discussion on this direction will be reported~\cite{ip}.

To apply our discussion to the quantization of SFT,
it may be convenient to lift our problem to the framework of BRST quantization where the physical state condition is written in  a form of one equation $Q \ket{\phi} =0$ and the residual gauge symmetry is represented by exact states $Q \ket{\chi}$ as $\ket{\phi} \sim \ket{\phi} + Q \ket{\chi}$. 
Even in the case of BRST quantization, we can naturally prove the corresponding statement as our theorem~1 itself and obtain the same result ${\cal H}_{\tilde{\epsilon}} \sim {\cal H}_{\rm obs}$, though in this case we have to impose appropriate conditions in the total state space including ghost states as additional gauge conditions. 
Actually, in ref.\cite{Asano:2000fp,Asano:2003qb}, BRST quantization of string theory on curved background represented by the CFT of the form $(c^0=1,h^0<0) \otimes (c^K=25,h^K>0)$ 
was considered and the claim that there were no negative-norm states in the observable Hilbert space was made. 
The logic used there was that the states with ghosts ($b_{-n},c_{-n}$) or time-like states ($\alpha_{-n}^0$) can decouple from observable Hilbert space.
Our present work for $\beta=0$ corresponds to giving explicit representation 
of the corresponding observable Hilbert space (without $b_{-n}$, $c_{-n}$ and $\alpha_{-n}^0$)
that had not been explicitly specified in \cite{Asano:2000fp,Asano:2003qb}. 
Furthermore, to proceed our discussion, we would like to find out whether the possible additional gauge conditions are expressed in simpler forms in terms of BRST quantization.

As stated in the introduction, our analysis is a first step toward a way of canonically quantizing SFT in the temporal gauge where the difficulty associated with the time-like nonlocality may be avoided. We may, however, learn from the analysis in the main section about the light-like gauge fixing of the covariant SFT as well. As is shown, the DDF states are the representation of physical states with the light-like gauge fixing condition. This means that the modes of the string field in this gauge will be expanded by the DDF states, so that the field in each mode has only transverse polarization. As far as the author's knowledge is concerned, there is no literature which derives the light-cone SFT by appropriately fixing the gauge in the covariant SFT. The detailed analysis of these issues will be reported elsewhere.

\section*{Acknowledgements}
%The work of M.A.~is supported in part by the Grants-in-Aid for Scientific Research from the Ministry of Education, Culture, Sports, Science and Technology (MEXT).
%That of M.K.~is supported in part by the Grants-in-Aid for Scientific Research from the Japan Society for the Promotion of Science (JSPS) and from the MEXT. 
%That of M.N.~is supported in part by the Grant-in-Aid for Scientific Research (13135224) from the MEXT.

The work is supported in part by the Grants-in-Aid for Scientific Research (17740142~[M.A.], 13135205 and 16340067~[M.K.], 13135224~[M.N.]) from the Ministry of Education, Culture, Sports, Science and Technology (MEXT) and from the Japan Society for the Promotion of Science (JSPS).

%%%%%%%%%%%%%%%%%%%%%%%%%%%%
\appendix 
\def\thesection{Appendix~\Alph{section}}
%%%%%%
\section{Proof of lemma~\ref{lemma1}} 
\setcounter{equation}{0}
\renewcommand{\theequation}{\Alph{section}.\arabic{equation}}
%%%
First, note that for each $\ket{f_{\beta};\,p} \in {\cal F}_{\beta} (p) $, a set of states  
\begin{equation}
L_{-n_1}^{\Sigma_\beta} \cdots L_{-n_r}^{\Sigma_\beta}
L^{t_\beta}_{-m_1} \cdots L^{t_\beta}_{-m_q} 
\ket{f_{\beta};\,p }
\label{eq:repHLLT}
\end{equation}
is equivalent to the set of states (\ref{eq:statebase1}) as a linear space since $L_{-n}=L_{-n}^{t_\beta}+ L_{-n}^{\Sigma_\beta}$.
Thus, it is sufficient to prove that the states (\ref{eq:repHLLT}) for all $\ket{f_{\beta};\,p} \in {\cal F}_{\beta} (p) $ are linearly independent and span a basis of ${\cal H}(p)$ if $p^{t_\beta}\ne 0$.

Recall that the total state space is divided into 
time-like $c=1$ and space-like $c=25$ part: 
$
{\cal H}(p) = {\cal H}_{t_\beta}(p^{t_{\beta}}) 
\otimes {\cal H}_{\Sigma_\beta}({p^{i_\beta}}) 
$.

For $c=1$ part, ${{\cal H}_{t_\beta}(p^{t_{\beta}})}$
can be represented by Verma module ${\cal V}(c=1,h^0)$
with highest weight $h^{0}= -\frac{1}{2} (p^{t_\beta})^2 $ since 
we know that ${\cal V}(c=1,h^0)$ is non-degenerate for $h^0<0$ from Kac's determinant
formula, i.e.,
\begin{equation}
{{\cal H}_{t_\beta}(p^{t_{\beta}})}
=  
\{ 
L^{t_\beta}_{-m_1} \cdots L^{t_\beta}_{-m_q} 
\ket{0,\,p^{t_\beta}}
\}
.
\end{equation}

For $c=25$ part, we would like to show 
that the space ${\cal H}_{\Sigma_\beta}({p^{i_\beta}})$ 
is spanned by the set of states
\begin{equation}
L^{\Sigma_\beta}_{-n_1} \cdots L^{\Sigma_\beta}_{-n_r} 
\ket{f_{\beta},p^{i_\beta}}
\, 
(\equiv
\ket{\lambda_{\Sigma n_i}\!= \{n_1,\cdots,n_r \}, f_{\beta} })
\label{eq:vermaf25} 
\end{equation}
with all $\ket{f_{\beta},p^{i_\beta}} \in {\cal F}_{\beta} (p^{i_\beta})$.
Note that the set of above states (\ref{eq:vermaf25})  forms the 
Verma module ${\cal V}(c\!=\!25,h)$ with $h=M+\frac{1}{2} (p^{i_\beta})^2$ for each $\ket{f_{\beta},p^{i_\beta}}$. 
%Each ${\cal V}(c\!=\!25,h>0)$ is unitary.
Dividing with each level $N$, the equation we would like to show 
is 
\begin{equation}
{\cal H}_{\Sigma_\beta}^{(N)} (p^{i_\beta})
= \bigoplus_{M=0}^{N}
\Big\{
\ket{\lambda_{N-M}, f_{\beta}^{(M)} }
\, \Big| \,
\ket{f_{\beta}^{(M)},p^{i_\beta} } \in {\cal F}_{\beta}^{(M)} ({p^{i_\beta}}) 
\Big\}
\label{eq:decompI}
%\\
%&\equiv& \sum_{N \ge 0} {\cal H}_{\Sigma_\beta}^{(N)}({p^{i}})
.
\end{equation}

We use the induction on $N$ to show eq.(\ref{eq:decompI}). 
For $N=0$, the equation is true trivially
since ${\cal H}_{\Sigma_\beta}^{(0)}({p^{i_\beta}}) = \{ \ket{0,\,p^{i_\beta}} \}$
and $\ket{0,\,p^{i_\beta}} \in {\cal F}_{\beta}^{(0)}(p^{i_\beta})$.
Then we suppose that the equation holds for level less than $N$ 
and consider the states at level $N$. 
We represent a state in ${\cal H}_{\Sigma_\beta}^{(N)}({p^{i_\beta}})$ as 
\begin{equation}
\ket{\psi_N}=\ket{g_N}+\ket{o_N},
\qquad \ket{g_N} \in {\cal G}^{(N)},
\qquad \ket{o_N} \in {\cal O}^{(N)}.
\end{equation}
Here ${\cal G}^{(N)}$ is generated by the states of the form 
$\ket{\lambda_{N-M}, f_{\beta}^{(M)} }$ 
with $M<N$:
\begin{equation}
{\cal G}^{(N)} =
\bigoplus_{M=0}^{N-1}
\Big\{
\ket{\lambda_{N-M}, f_{\beta}^{(M)} }
%\, \Big| \,
%f_{\beta}^M \in {\cal F}_{\beta}^{(M)}, \{\lambda_{N-M} \}
\Big\}
\end{equation}
and ${\cal O}^{(N)}$ is the complement of ${\cal G}^{(N)}$ in 
${\cal H}^{(N)}_{\Sigma_\beta}(p^{i_\beta})$.
A state  \ket{g_N} has non-trivial inner products only within ${\cal G}^{(N)}$ and ${\cal G}^{(N)}$ is non-degenerate 
since ${\cal V}(c=25,h>0)$ does. 
Thus, ${\cal O}^{(N)}$ is orthogonal to ${\cal G}^{(N)}$:
\begin{equation}
{\cal H}^{(N)}_{\Sigma_\beta}(p^{i_\beta}) =  {\cal G}^{(N)}  \oplus  {\cal O}^{(N)} .
\end{equation}
Consider a state $ L_{-m}^{\Sigma_\beta} \ket{\psi_{N-m}} \in  {\cal G}^{(N)}$
with $\ket{\psi_{N-m}} \in {\cal H}^{(N-m)}_{\Sigma_\beta}(p^{i_\beta})$
$(m \ge 1)$. 
Since ${\cal O}^{(N)}$ is orthogonal to ${\cal G}^{(N)}$,
\begin{equation} 
(L_{-m}^{\Sigma_\beta} \ket{\psi_{N-m} } )^\dagger \ket{o_N} = 
\bra{\psi_{N-m}} L_{m}^{\Sigma_\beta} \ket{o_N}
= 0
\end{equation}
for any state $\ket{o_N} \in {\cal O}^{(N)}$.
From the fact that $L_{m}^{\Sigma_\beta} \ket{o_N} \in {\cal H}^{(N-m)}_{\Sigma_\beta}(p^{i_\beta})$
and that ${\cal H}^{(N-m)}_{\Sigma_\beta}(p^{i_\beta})$ is non-degenerate, we must conclude that
\begin{equation}
 L_{m}^{\Sigma_\beta} \ket{o_N} =0 \quad (m\ge 1),
\end{equation}
which indicates that $\ket{o_N}$ is nothing 
but an element of ${\cal F}_{\beta}^{(N)}$.
Thus,
\begin{equation}
 {\cal H}^{(N)}_{\Sigma_\beta}(p^{i_\beta}) =  {\cal G}^{(N)}  \oplus {\cal F}_{\beta}^{(N)}
\end{equation}
and this means that the equation (\ref{eq:decompI}) holds for $N$.

Combining with the result for $c=1$ part, we have completed the 
proof of lemma~\ref{lemma1}.
\qed

%%%%%%%%%%%%%%%%%%%%%%%%%%%%%%%%%%%%%%%%%%%%%%

\end{document}